\documentclass[12pt]{article}
\usepackage{graphicx}
\usepackage{amsmath}
\usepackage{amssymb}
\usepackage{authblk}
\usepackage[left=0.9in,right=0.9in,top=0.8in,bottom=0.8in]{geometry}
\usepackage{setspace}
\usepackage{float}
\usepackage[super,sort&compress,comma]{natbib}
\usepackage[version=4]{mhchem} 
\usepackage{graphicx} 
\usepackage[font={normalsize}]{caption} 
\usepackage[labelfont=bf]{caption}
\usepackage{amsmath, amssymb}
\usepackage{natbib}
\usepackage{booktabs}
\usepackage{array}
\usepackage{subcaption}
\usepackage{multirow}
\usepackage{soul}
\usepackage{float}
\usepackage{siunitx}
\usepackage{tabularx}
\usepackage{ulem}
\usepackage{hyperref} 
\usepackage{xcolor}
\usepackage[labelfont=bf]{caption}
\usepackage[font={small}]{caption} 

\newcommand*\edit[1]{{\color{blue} #1}}

\begin{document}

\title{\textbf{Teachers that teach the irrelevant: Pre-training machine learned interaction potentials with classical force fields for robust molecular dynamics simulations}}

\author{Eric C.-Y. Yuan$^{1,4}$ and Teresa Head-Gordon$^{1-4}$}
\date{}
\maketitle

\begin{center}
\vspace{-24pt}
$^1$Kenneth S. Pitzer Theory Center and Department of Chemistry, $^2$Department of Bioengineering, $^3$Department of Chemical and Biomolecular Engineering,  University of California, Berkeley, CA, 94720 USA\\
$^4$Chemical Sciences Division, Lawrence Berkeley National Laboratory, Berkeley, CA, 94720 USA

corresponding author: thg@berkeley.edu
\end{center}

\begin{abstract}
\noindent
Machine learned interaction potentials (MLIPs) have become a critical component of large-scale, high-quality simulations for a range of chemical and biochemical systems. Yet, despite their in-distribution accuracy, molecular dynamics simulations using MLIPs exhibit numerical instabilities due to underlying data insufficiencies when encountering new regions of the potential energy surface. Here we propose a pre-training learning scheme that uses low-quality, practically free, single-molecule non-reactive force field data while all intermolecular interactions and reactive properties are learned at a fine-tuning stage with a small amount of computationally more expensive labels. We show that the force field pre-training approach followed by data efficient ab initio fine tuning allows for stable and accurate molecular dynamics and metadynamics simulations of gas phase molecules, liquid water, and hydrogen combustion reactions compared to models trained from scratch.

\end{abstract}

\section{Introduction}
\vspace{-3mm}
\noindent
The ongoing development of machine learning interatomic potentials (MLIPs) has transformed the field of computational chemistry.\cite{Behler2015,Drautz2019,Gilmer2017,Lubbers2018,Batzner2021,Haghighatlari2022,batatia2025,Yuan2025,wood2025} By training deep learning models on high quality ab initio calculations for energy and forces, explorations on the potential energy surface (PES) using molecular dynamics (MD) simulations can be several orders of magnitude faster while obtaining good chemical accuracy compared to ab initio molecular dynamics (AIMD). Recent advancements in universal potentials, which are MLIPs trained on millions of configurations across a range of molecular systems at a targeted level of typically Density Functional Theory (DFT), demonstrates that increasing amounts of energy and force data can be immensely helpful for transferability to related chemical systems.\cite{Batatia2023,Kovacs2025,Anstine2025,levine2025,batatia2025,wood2025}. 

While MLIPs demonstrate high accuracy for in-distribution (ID) samples, they struggle in extrapolating to out-of-distribution (OOD) test examples given the limited amount of ab initio training data across all regions of the PES.\cite{Behler2015,Smith2019,Yang2022,Guan2023} Hence during a MD trajectory the MLIP encounters ``holes'' in the PES landscape such that MLIPs cannot run stable simulations on long timescales, a significant failure that has been demonstrated on various chemical systems and MLIP architectures. \cite{Fu2022,Stocker2022,Vita2023,Morrow2023,Guan2023,Bihani2024} The most common strategy to learn OOD samples when encountering new regions on the PES is simulation-based test-time adaptation using active learning.\cite{Smith2018,Schran2020,Vandermause2020,Schran2021,Lin2021,Yang2022,Kulichenko2023,Guan2023} The purpose of the active learning stage during the MD simulation is to detect OOD errors, to select batches of frames to be labeled using the original ab initio data source, and the MLIP is then retrained on the expanded dataset. Multiple cycles of new labeling and retraining are usually necessary, typically involving tens to hundreds of iterations, and is a hidden cost that significantly decreases the overall computational efficiency gains made by simulating with a MLIP force field compared to more robust methods such as AIMD. 

The chief reason for such errors is the lack of enough high-quality labeled data that can comprehensively cover the entire conformation space. This is because most MLIPs are trained on low energy and/or metastable transition states due to their chemical significance, whereas unphysical states are often ignored in data acquisition because they are deemed chemically irrelevant or "poison" the training.\cite{Kulichenko2023,Guan2023} And yet in the high-dimensional space of a PES most regions are in fact chemically irrelevant and are thus undersampled, whereas equilibrium and meta-equilibrium states are relatively sparse but are oversampled during the data acquisition process for training. While the MD simulation stability is apparently a data issue, \textit{a priori} construction of a set of training and test samples to more comprehensively cover all regions of the PES has proven to be challenging.\cite{Stocker2022,Kulichenko2023,Guan2023} For example, both MD17 and MD22 datasets were created by running 500 K AIMD simulations, higher than the test temperature of 300 K when running molecular dynamics of the trained MLIP\cite{Chmiela2017,Chmiela2018,Chmiela2023}, along with careful design of data samples to maximize the coverage of the conformation space even with higher energy data.\cite{Cheng2019} Another strategy is to explore high energy regions within the lower manifold of collective variables as a systematic way to collect high energy and force samples.\cite{Yang2022,Kulichenko2023,Guan2023,Novelli2024} Another would be self-supervised training methods that align the simulated and reference distributions \cite{Raja2024,Cui2025,gardner2025} to address the problem of undersampled data, but overall these are still active learning processes that require ongoing accumulation of expensive ab initio labels for the OOD examples. As shown by Guan et al., at some point it is equally or even more efficient to simply take the energy and force from the original ab initio data source directly to advance the trajectory without retraining\cite{Guan2023}. 

In this work we propose a fully data-driven approach to avoid active learning or any kind of test-time adaption, by separating the MLIP training into a pre-training (PT) stage and a fine-tuning (FT) stage. The central idea is that the PT phase uses a large amount of classical force field (FF) data based on molecules or molecular fragments that sample high energy states, relegating the FT phase to use only a small amount of high-quality ab initio data for the physically or chemically relevant states, such as equilibrium conformations, reactants, products, and transition states. Unlike transfer learning where the goal is to learn higher quality labels with fewer training data, for example, from semi-empirical methods, lower rung DFT functionals, or Møller–Plesset perturbation theory to hybrid functionals or gold standard coupled cluster level data\cite{Smith2019,Zheng2021,Kser2022,Chen2023,Zaverkin2023,Kirschbaum2024,Khazieva2024}, the goal here is to pre-train in order to precondition the MLIP to comprehensively smooth the PES in all regions of phase space. In contrast to data augmentation using Morse potentials for chemical reactivity \cite{Luan2025} or use of empirical FFs or kernel-based MLIPs in pre-training that aim to suitably capture chemically relevant conformations \cite{Shui2022,Wang2024,Gardner2024,Gardner2023}, the crucial distinction of our FF pre-training (FFPT) followed by fine-tuning (FFPT-FT) strategy is not to improve the ID accuracy (although we find it often does) but to increase the OOD robustness where high-quality labels are not available. Furthermore, our energy and force labels come from single-molecule, non-reactive FFs, which are practically free in cost but still capture important physical behaviors. Like distillation approaches\cite{Alkhulaifi2021} by which broader labeled data learned from one model (the teacher) is transferred to another more specialized and typically smaller model (the student), in our case the teacher provides nothing but poor and/or chemically irrelevant examples while the student specializes to chemical relevancy for better performance across diverse applications. The FFPT-FT strategy demonstrates qualitative improvement on various representative chemical systems, including stable MD simulations of individual organic molecules, condensed phases illustrated with bulk water, and chemical reactivity of reaction channels in hydrogen combustion, showing the generality of the approach.
\vspace{-3mm}

\section{Results}
\vspace{-3mm}

\begin{figure}[tb!]
\centering
\includegraphics[width=0.9\textwidth]{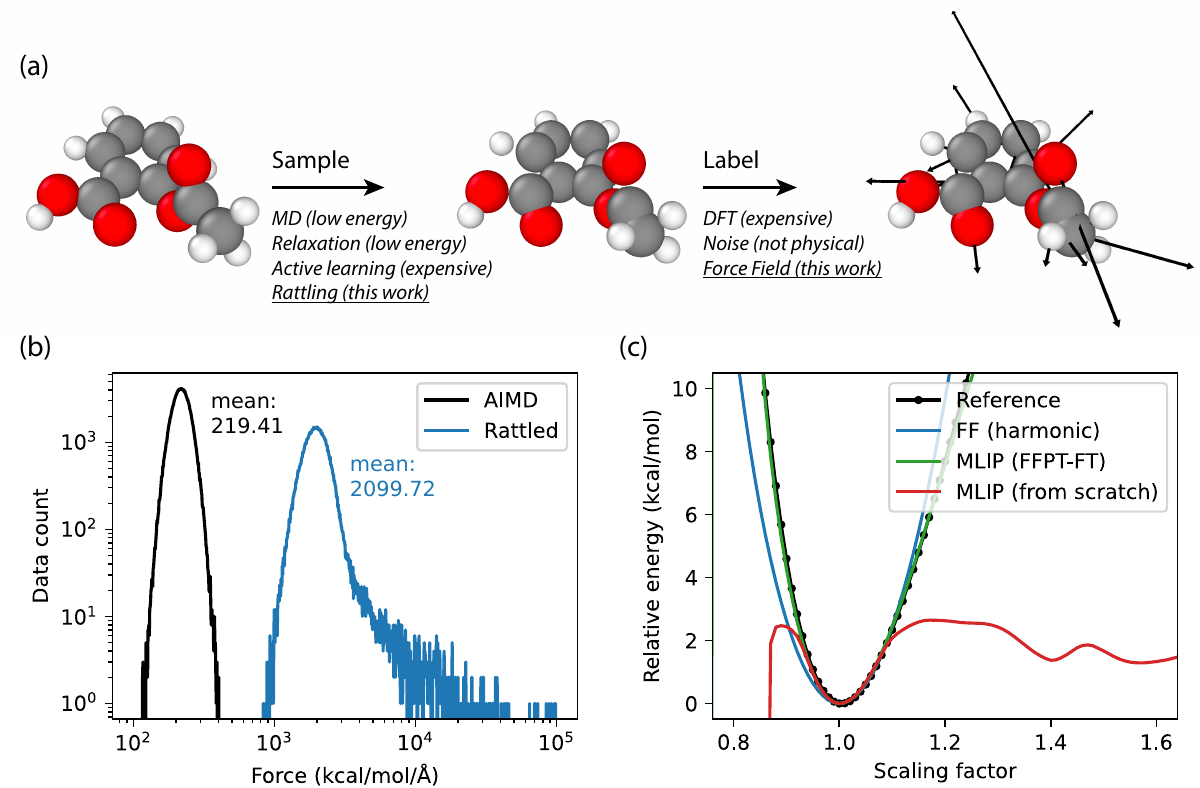}
    \caption{\textit{Force field strategy of sampling high energy and unphysical data for pre-training a MLIP with subsequent fine-tuning.} (a) The general workflow for chemical dataset construction can be divided into sampling and labeling. We use rattling to systematically sample high energy conformations, as well as using physics-based FFs to label the data to ensure data coverage in unphysical regions. (b) Compared to accumulating frames from AIMD simulations at high temperatures (black), rattling (blue) generates more uncorrelated structures, effectively corresponding to running simulations at 45,000 K. (c) The PES from an MLIP trained on chemically relevant data (red) tends to be rough and incomplete and on a lower energy scale, leading to simulation instabilities. The PES from FFPT (blue) has the correct limiting behavior despite its  lower accuracy, but can be fine-tuned for a more complete PES with high quality data (green). The potential energies are normalized per vibrational degree of freedom.}
    \label{fig:data}
\end{figure}

Figure \ref{fig:data}a defines the general procedure for our chemical dataset construction using rattling to systematically sample high energy conformations, and is distinguished from past work by using simple FFs to efficiently label data with physically reasonable, though not accurate, energies and forces\cite{}. As opposed to AIMD simulations at elevated temperatures (black), we use rattling (blue) to generate completely uncorrelated structures (Figure \ref{fig:data}b). Since the purpose of the PT stage is to pre-condition the MLIP for PES smoothness and limiting behaviors, we do not specifically sample important low energy parts of the chemical space, as typically desired using MD trajectories, relaxation trajectories, or conformation scans, nor do we need expensive cycles of active learning to explore the PES. Rattling has been used in more recent data generations efforts\cite{BarrosoLuque2024,wood2025}, and more broadly can be considered as a type of noised input. Such an added noise has been shown to be a type of harmonic FF,\cite{Zaidi2022} though not parametrized on a real PES and often violating the translational and rotational symmetry of a molecule, but this is not an issue for the PT phase. 

As seen in Figure \ref{fig:data}c an MLIP trained from scratch (red) tends to be too rugged and lacking a more complete energy scale for molecular interactions (i.e. softening\cite{Kreiman2025,Yuan2025}), which would lead to simulation errors and instabilities. The `holes'' on the PES are particularly concerning, where the atoms crash into (left-most region) or are torn apart from (right-most region) each other but are predicted to be of low energy. However, these unphysical configurations have extremely high energies in AIMD and thus are poorly sampled in the training data, and are states that are hard to capture through detailed by-hand construction. Instead, a PES derived from FFPT (blue) has the correct limiting behaviors for high energy states despite its lower accuracy, and serves as a generic PT model that will ensure MD stability having conditioned on PES smoothness everywhere. The FFPT model can now be fine-tuned with the high quality or chemically relevant, but more sparsely available ab initio data source, such that the FFPT-FT combination (green) will yield MD simulation stability and accurate prediction of the reference DFT data. 

A reasonable question is whether such a multi-fidelity training scheme would compromise the prediction accuracy of the final model. Indeed, mixed data quality should generally be avoided, and that's why we separated the use of low- and high-quality data into PT and FT stages, respectively. By separating the training into multiple stages, the ID accuracy is determined by the last training step, which is fine-tuned using the high-quality DFT data. In alignment with transfer learning protocols \cite{Shui2022,Wang2024,Gardner2024}, and in contrast to data augmentation approaches \cite{Luan2025}, the data from different levels of theory is never mixed together. In subsequent sections we investigate how well a cheap FF for single molecules can serve as a pre-trained model that ensures stable MD trajectories for gas phase molecules and condensed phase periodic liquid systems, as well as for chemical reactivity using metadynamics for hydrogen combustion reactions.

\subsection{Force field pre-training strategy for small molecules}

\begin{figure}[tb!]
\centering
\includegraphics[width=0.9\textwidth]{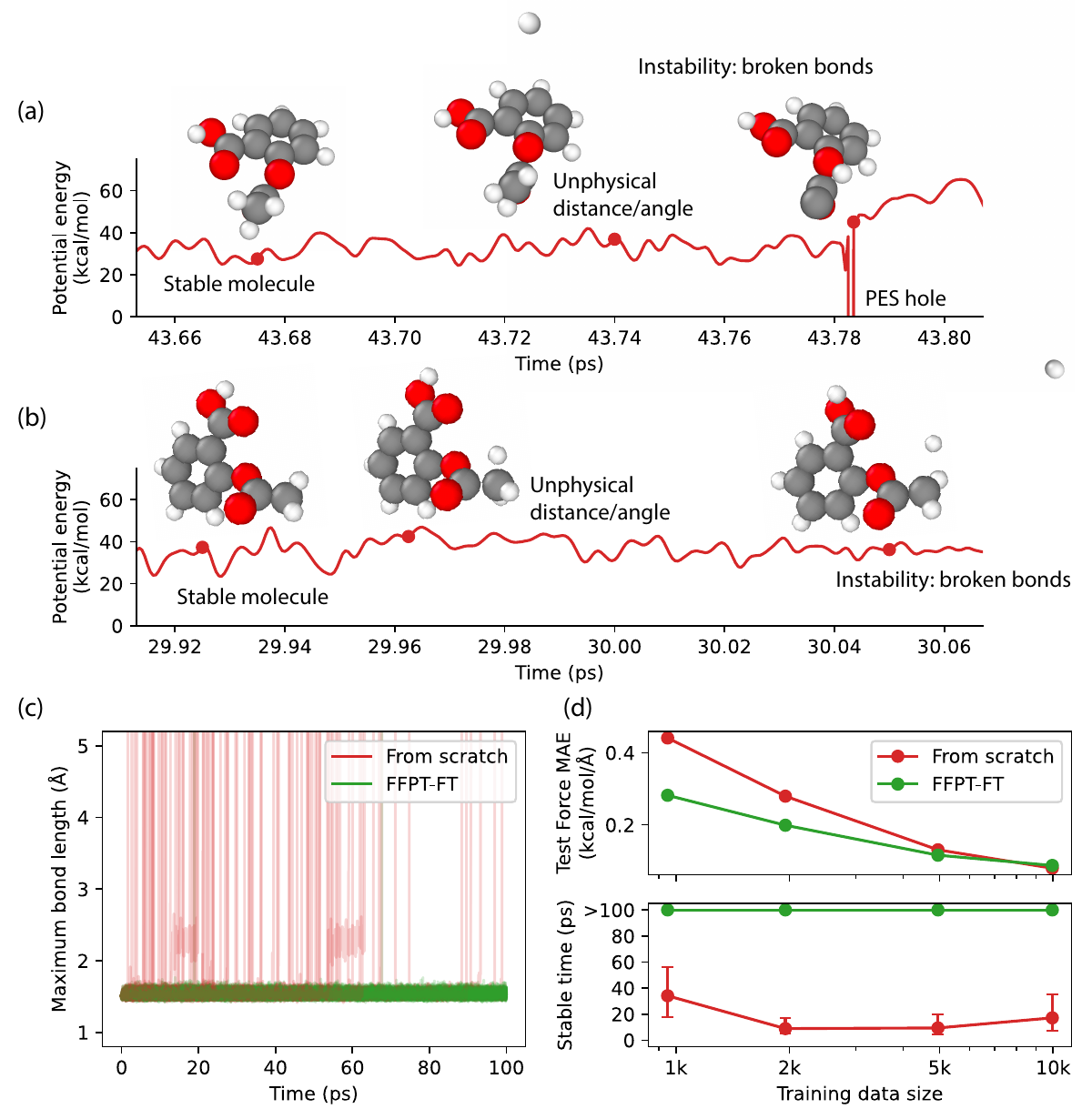}
    \caption{\textit{MD simulation stability improved by FFPT for aspirin.} (a,b) MD failures can occur with or without hitting a hole on the PES. (c) FFPT greatly improves the MD stability compared to an MLIP trained from scratch. (d) The stability improvement does not come from the ID accuracy. Even with more training data and lower error, the MD stability does not improve correspondingly. }
    \label{fig:aspirin}
\end{figure}

We first test our FFPT-FT approach on a seemingly trivial system: a single non-reactive organic molecule in vacuo. Figures \ref{fig:aspirin}a and b show two failure modes of the MD simulation for MLIPs trained from scratch. In the first case, the HCH angle becomes unphysically small, and the two Hs eventually crash into each other and fly apart due to the large force, and in the second case for which the C-H distance becomes unphysically long and the H gradually drifts away without any energy/force consequences. The common theme is that the simulation using the MLIP runs into an OOD region of the PES, where, without further knowledge about it, the MLIP predicts an average low energy and allows the unphysical state to be visited. 

This is a direct result from the wrong limiting behaviors of the PES in Figure \ref{fig:data}c, which can be easily fixed by FFPT. With its improved OOD limiting behaviors, the simulation instability is indeed resolved and without requiring any active learning. As demonstrated in Figure \ref{fig:aspirin}c, the MLIP trained from scratch showed frequent unphysical bond dissociation events within tens of picoseconds, which is not energetically possible at 500 K, whereas the FFPT-FT MLIP correctly describes the molecular stability. Consistent with literature results, this improvement is independent of the ID test error shown in Figure \ref{fig:aspirin}d.\cite{Fu2022} While the FFPT improves the test error by reducing overfitting, the MD stability does not significantly change with the test accuracy with more training data, showing that the MD simulation benefits from better OOD performance.

We also benchmarked our FFPT-FT approach against more established training protocols in this well-studied system to acquire a more comprehensive comparison. Delta learning approaches from a baseline FF such as the Ziegler-Biersack-Littmark (ZBL) potential\cite{Biersack1982} have been proposed to handle the limiting behavior of atoms and improve the simulation stability for MLIPs.\cite{Batatia2023} However, Supplementary Figure 1 shows that this is an ineffective approach since it is designed for extremely short range repulsion, and modifications over more physical ranges would simply interfere with the accurate regions of the PES. Active learning is  another popular approach,\cite{Smith2018} but not only is it unnecessarily expensive in both sampling and labeling as discussed previously, the newly labeled high-energy data inevitably dilutes the more chemically relevant original training set, significantly lowering the ID accuracy, as shown in Supplementary Figure 2. Hence our separation of the training stages essentially provides more robust results compared to these approaches. Finally, we also show in Supplementary Figure 3 that the choice of the pretraining FF is not critical, such that a simple Lennard Jones model is found to be sufficient as well.

\subsection{Force field pre-training strategy for liquid water}

\begin{figure}[tb!]
\centering
\includegraphics[width=0.9\textwidth]{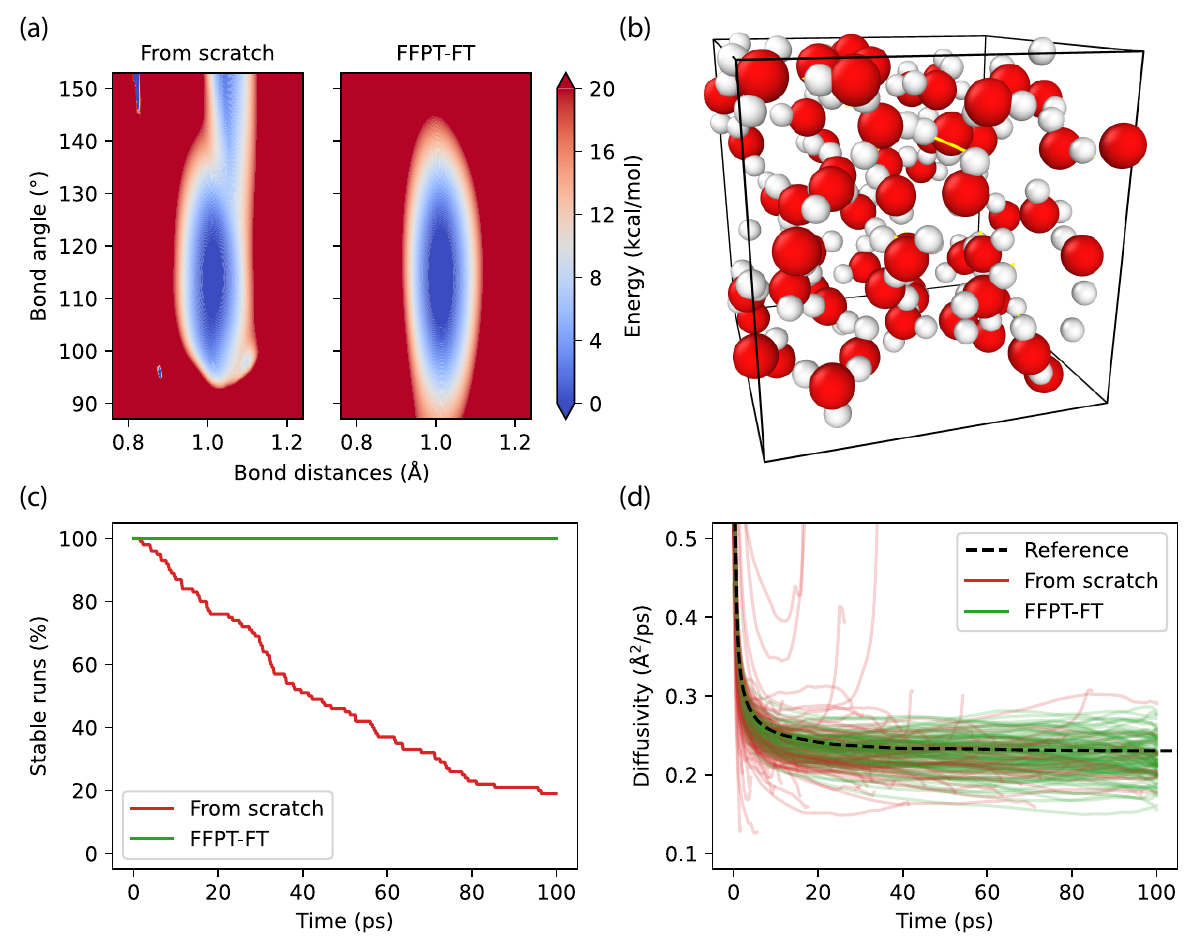}
    \caption{\textit{Bulk water simulation stability improved by monomer FF pre-training.} (a) The MLIP trained from scratch has holes in the PES unlike the FFPT-FT for the water monomer. (b) In the condensed phase simulation using the MLIP trained from scratch, water molecules can adopt a near-linear conformation which leads to collisions with neighboring waters. (c) By pre-training on a one-body FF and fine-tuning with bulk water data, the MD simulation is stable (green) unlike the MLIP trained from scratch (red). (d) A more stable simulation is critical for accurate evaluation of dynamical properties such as the diffusivity.}
    \label{fig:water}
\end{figure}

We next consider the FFPT performance for a periodic system compared to an MLIP developed from scratch for bulk water.\cite{Fu2022} But instead of pre-training an MLIP with FF data for a box of water, and fine-tuning the PES with ab initio data, we'd like to take a step further to pre-train the model with FF data for only monomers. This is because sampling a many-body PES is a non-trivial task by itself, where exhaustive enumeration of configurations of dimers, trimers, and tetramers becomes rapidly impractical for even advanced FF models such as MB-Pol\cite{Paesani2023}, Q-Aqua\cite{Bowman2022}, and CMM\cite{Heindel2024}. Furthermore, the non-bonded interaction energies are on a different energy scale compared to distortions of intramolecular bonds and angles. Hence we test the FFPT approach using a monomer FF, relegating learning all intermolecular interactions during the FT stage.

As shown in Figure \ref{fig:water}a, the MLIP trained from scratch has an unphysical but low energy ``hole'' on the monomer surface at $>150^\circ$ angle. During the liquid water simulation, one of the water molecules can fall into this near-linear configuration, thereby occupying a larger excluded volume and clashing with a nearby water molecule as seen in Figure \ref{fig:water}b. This steric overlap is a negative example, as has been shown on aspirin, and results in an unstable simulation that quickly fails within tens of picoseconds using the MLIP trained from scratch (Figure \ref{fig:water}c). Applying FFPT on the monomer surface removes this artifact as seen in Figure \ref{fig:water}a, and the FFPT-FT model remains perfectly stable over the simulation timescale of 100 ps. The MD stability is crucial for collection of physical and chemical properties, one of which is the diffusivity shown in Figure \ref{fig:water}d, for which the FFPT-FT model is able to accurately calculate the diffusion constant of water, unlike the MLIP trained from scratch.

\subsection{FFPT with non-reactive force fields for chemical reactions}
\label{sec:h2comb}

The most important use case of MLIPs in chemistry is arguably for reactive systems. Recently Guan and co-workers have created chemically relevant data sets using intrinsic reaction coordinates (IRCs), AIMD trajectories near transition states (TS), and normal mode sampling along IRC paths, for all 19 reaction channels for hydrogen combustion (see Supplementary Table S1).\cite{Guan2022} Even so, an MLIP model trained from scratch on this data was found to be incomplete, requiring active learning to capture the high-dimension and unphysical regions of the PES\cite{Guan2023} that were not accounted for in the original data.\cite{Guan2022} However, even after 50 rounds of active learning resulting in an additional 46,182 DFT energy and forces added into the dataset, the PES of the final trained model was still incomplete such that the MD was sporadically unstable.\cite{Guan2023} Instead of the expensive cost of data generation and retraining using active learning, it was found that a hybrid model that directly substituted the force from the original DFT data source to complete unstable MD steps performed well with no cost in computational efficiency.\cite{Guan2023} 

Here we consider whether we can avoid the deployment of active learning altogether using the FFPT-FT strategy for the case of hydrogen combustion. We pre-trained an MLIP using a non-reactive FF developed in Q-Force\cite{Sami2021} (see Methods) for the 8 reactant and product species \ce{O, H, H2, O2, OH, H2O2, H2O, HO2}. The FFPT model is expected to interpolate between the reactant (R) and product (P) energy minima, and the reactivity can be learned in the FT stage using the chemically relevant data from IRCs, AIMD trajectories, and normal mode sampling for all 19 reaction channels that preceded the active learning\cite{Guan2022}. We also create an MLIP trained from scratch from the same earlier set of hydrogen combustion data\cite{Guan2022}, and which includes no "negative" examples from active learning nor calls to the original DFT source itself as done in the hybrid model developed later by Guan and co-workers.\cite{Guan2023} 

\begin{figure}[tb!]
\centering
\includegraphics[width=0.9\textwidth]{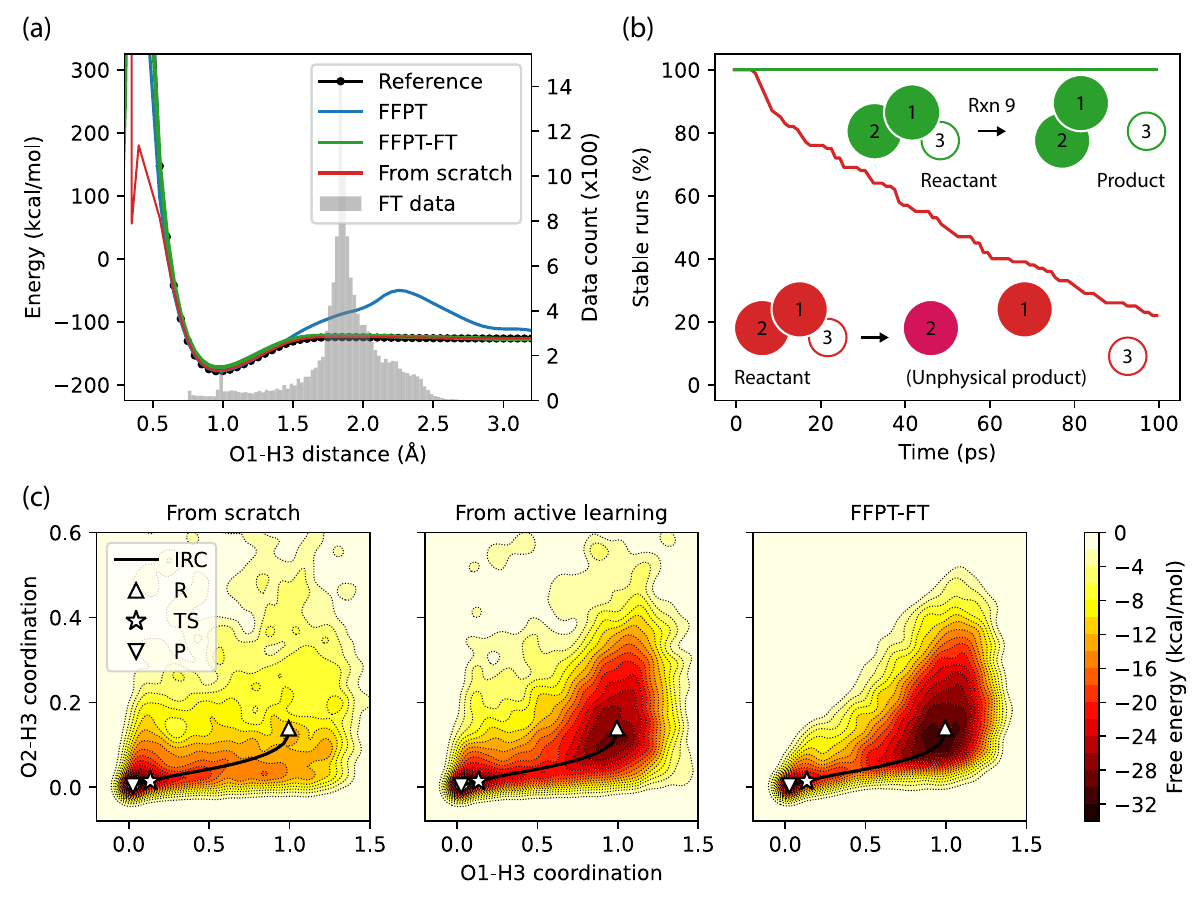}
    \caption{\textit{Hydrogen combustion reactions improved by non-reactive FFPT} illustrated using reaction 9 \ce{HO2 -> H + O2}. (a) When pre-trained on non-reactive FFs for reactant and products, the FFPT model can learn an effective interpolation over the course of reaction described by the O1-H3 order parameter (blue). While not quantitatively accurate, it can be accurately fine-tuned using high-quality positive examples from DFT (green). In contrast, a model trained from scratch showed catastrophic OOD behaviors in simulation at short-range (red) due to missing high energy data (gray). (b) The non-smooth PES leads to simulation failure by predicting unphysical products when running metadynamics with the MLIP from scratch (red) whereas the MD is stable using the FFPT-FT model (green) that yields correct products. (c) The unstable simulation trajectory of the MLIP from scratch reconstructs a free energy surface that does not resemble the ground truth, with an overstabilized product state with a large entropy component to the free energy. While improved with active learning, the FFPT-FT approach is superior due to complete MD stability over the simulation run.}
    \label{fig:h2comb}
\end{figure}

Figure \ref{fig:h2comb}a compares different MLIP models for reaction 9, \ce{HO2 -> H + O2}. The FFPT model has successfully learned to interpolate between the non-reactive FFs for the reactant \ce{HO2} (R) and the product \ce{H + O2} (P), and a transition state (TS) is predicted. The exact location of this TS and its energy are not accurate compared to the reference DFT, but the overall PES is smooth such that the FT process can effectively deal with the ID accuracy. As a result, the FFPT-FT model shows the same high accuracy as the model trained from scratch in the chemically relevant regions, while also improving the OOD performance by avoiding the unphysical clashing at short O...H distances where no DFT data is sampled.

To address an even more difficult use case we consider the construction of the free energy surface (FES) using metadynamics\cite{Bussi_Laio_2020}, in which the molecules are constantly driven to react within a well-defined set of collective coordinates, and for which other high-energy configurations are also frequently visited. Even though the original hydrogen combustion dataset\cite{Guan2022} contains chemically relevant labeled data near the TS of the IRC, the data generation failed to avoid \ce{O2} bond breaking that creates the MD instability issue seen in Figure \ref{fig:h2comb}b for the MLIP trained from scratch. This results in a manifest problem in the FES as seen in Figure \ref{fig:h2comb}c because the OOD PES errors arise from assuming \ce{O2} bond breaking is low energy, leading to a large but false entropy component in the product well. While the MLIP model trained on the active learning dataset\cite{Guan2023} covers a wider region of the PES and improves the FES as a result, it still inherits some of the unphysical product features observed for the from-scratch model due to MD instability. The FFPT-FT model, on the other hand, is able to run long, stable metadynamics simulations without any active learning or extra DFT calculations involved, and the FES is devoid of false entropy stabilization that in turn now deepens the reactant well. 

In addition to reaction 9, we found that reaction 1 \ce{OH + O -> H + O2} also showed similar problems of unphysical product states that were better addressed by the FFPT-FT approach as seen in Supplementary Figure 4. While the extensive active learning phase resulted in stable metadynamics for generating converged FESs for the other reaction channels (Supplementary Figure 5), the FFPT-FT approach converges to the same FES without any need for the 50 rounds of active learning using expensive labeled data, and illustrates the efficiency of the FF approach.

\section{Discussion and Conclusion}
The core idea of this work is that MLIP instability is a data issue such that we can apply better data-centric solutions for more stable MD simulations. A transfer learning approach is applied which separates the training into PT and FT stages, the former of which does not require high-quality data. We demonstrated that the PT data can be incommensurate with the downstream FT data, in our case the use of classical single-molecule non-reactive FFs, which imbue the MLIP with physical robustness despite its moderate accuracy and even its lack of chemical relevance. Hence an almost infinitely large amount of low-quality labeled data can be generated, and high-energy, high-force region can be intentionally sampled to improve the OOD performance, instead of being filtered out due to the fear of data poisoning at the energy scale of chemical relevance. We showed that intramolecular and intermolecular interactions and reactivities can be learned during the FT stage with a small, high-quality ab initio dataset that only needs to cover the PES regions of chemical interest.

We envision the FFPT-FT approach can be almost trivially applied to various systems, including small and large molecules, condensed phases, materials, solutions, interfaces, chemical reactions, catalysis, and beyond. \edit{A relevant application of our method is for model distillation, which can be viewed as training a small model from scratch using labels from a large, more expensive MLIP. Pre-training the student model with cheap FF data could still be useful given the growing size and thus the slower inference speed of the state-of-the-art foundation models. Even if one considers the labeling cost from the teacher model to be negligible, the separation of training stages and the sampling method proposed in this work remain applicable in this case.} 

Though demonstrated with the equivariant NewtonNet MPNN\cite{Haghighatlari2022}, the proposed method is notably agnostic to the choice of MLIP architectures, as the simulation stability is a universal problem to most MLIPs, if not all. The method is independent of the choice of the FT method too. The FT strategy we applied in this work is a simple supervised training on a single set of data for the entire model, and one could imagine to exploit more advanced FT strategies such as multi-fidelity/multi-objective training\cite{wood2025,messerly2025} or even further transfer learning over different levels of ab initio datasets\cite{Hoffman2023}. We also leave the more detailed investigations of the transfer learning strategies like low-rank approximation\cite{mcgrath2025} to future studies.

We must emphasize this work does not diminish the importance of high-quality DFT datasets. The current universal potentials, powerful as they already are, are MLIPs trained on one to a hundred million energy/force labeled configurations using good to excellent DFT functionals\cite{levine2025,Anstine2025,Eastman2023,Yuan2025}, and most recently the OMol25 data set utilized rattling to generate high energy and high force structures. On the other hand, the dataset size can go to one to five billion by giving up DFT labels \cite{Irwin2020,Tingle2023,Ji2024} and hundreds of billions by further giving up 3D structures \cite{Ruddigkeit2012}. We believe our FFPT approach based on practically free FF labels approaches complete chemical coverage, and the expensive, high-quality DFT data can be used for the FT phase of the model for chemical accuracy.

At the same time our current FFPT setup doesn't solve all the MLIP issues. While we include unphysical conformations by rattled sampling and FF labeling, we do not include unphysical chemical compositions or even valid chemical compositions that are unanticipated such as the appearance of the high energy hydronium ion for hydrogen combustion (Supplementary Figure 6). Even though some contrastive learning approaches apply node deletion or replacement to augment the composition space\cite{Fang2023}, it's unclear to us whether they could be applied in our case. Since over-coordination means more nodes have to be added to the molecular graph, new augmentation schemes may have to be developed. Finally, we do expect that for models and FT data that incorporate multiple charge/spin states could significantly alleviate this issue\cite{levine2025,Yuan2025}.

Finally, the FFPT as an independent PT module therefore leads us to another possibility for defining a chemical foundation model. If the FF parameters can be pre-assigned, generating FF labels is so cheap that it might be able to be achieved on-the-fly, at a similar cost to self-supervised learning. Certainly, more work has to be done to understand whether an FFPT model can be useful in properties beyond energies and forces. Yet, compared to models trained on denoising tasks \cite{Zaidi2022,Wang2023,Jiao2022,Liu2022,Zhou2018,Ji2024}, we believe a model that resembles an approximate but physically meaningful PES that respects chemical bonding may be a contender as a foundation model for other interesting chemical properties.

\section{Methods}

\noindent
\textbf{Pre-trained data from force fields}. Only isolated atoms, molecules, and molecular fragments are considered in the generation of data for the MLIP, i.e. one aspirin molecule, a water monomer, and isolated reactants and products for the hydrogen combustion model (H, O, \ce{H2}, \ce{O2}, OH, \ce{H2O}, \ce{HO2}, and \ce{H2O2}). We generated the pre-training data using GAFF-2.11 for aspirin, and for the water monomer using the flexible version of TIP3P water model.\cite{Jorgensen1983} For the hydrogen combustion study, we parametrized a non-reactive force field for all chemical species in their relaxed structures using Q-Force\cite{Sami2021} with the same functional and basis set as the original hydrogen combustion dataset.\cite{Guan2022} Parameters for the Q-Force model are provided in Table \ref{tab:ff}.\\

\noindent
\textbf{Pre-trained data sampling}. 100,000 uncorrelated frames were generated around the force field relaxed geometries. To sample high-energy unphysical geometries, we added Gaussian random noise ${\epsilon}$ with a scale of 0.2 \AA to the atomic positions. Typically in denoising tasks, the original data is scaled down to $\tilde{\ce{x}}=\sqrt{1-\sigma^2}\ce{x}+\sigma\ce{\epsilon}$ preserve the variance of the data,\cite{Ho2020} but we chose not to scale the original positions so that $\tilde{\ce{x}}=\ce{x}+\sigma\ce{\epsilon}$, similar to related works.\cite{Zaidi2022} This scheme tends to expand the molecule, which resembles a collection of atoms at high temperature, rather than preserving the rough atomic density. The ensemble is more realistic to the OOD data encountered in MD simulations, yielding a better empirical performance. \\

\noindent
\textbf{Fine-tuning data}. We derived the aspirin data from the MD17 dataset\cite{Chmiela2017} For each molecule, we randomly sample 950 to 9,950 configurations for training and 50 for validation from the MD17 database. We randomly sample 10,000 configurations from the rest of the
data for force error evaluation. The bulk water data is from the reference literature\cite{Fu2022}, randomly sampled 950 frames for training and 50 for validation. The hydrogen combustion data includes only the original dataset\cite{Guan2022} for the from-scratch model, and both the original and the dilation/active-learning dataset\cite{Guan2023} for the from-active-learning model, randomly sampled 10\% for training and 1000 for validation.\\

\noindent
\textbf{Model pre-training protocol}. All models applied the equivariant NewtonNet architecture with 128 node features, 20 radial basis, 5 \AA distance cutoff, and SiLU activation.\cite{Haghighatlari2022} A single energy head and the corresponding gradient force head were used for prediction, despite conflicting examples might be given. The Adam optimizer\cite{Kingma2014} was used with decay rate of 0.7 on plateau with 50 epoch patience until 1\% of the initial learning rate. The initial learning rate is set to $10^{-3}$ for training from scratch and $10^{-4}$ for pretraining and finetuning. We took the energy weight of 1 and force weight of 20 in the loss function to put more emphasis on forces for derivative properties.\\

\begin{table}[H]
    \centering
    \caption{\textbf{Force field parameters for hydrogen combustion.} The force field terms used in Q-Force include $E_\text{bond} = D\left(1-e^{-k_r(r-r_0)/2D}\right)^2 - D$; $\quad E_\text{angle} = \frac{k_\theta}{2} (\theta - \theta_0)^2$; $\quad  E_\text{dihedral} = k_\phi (1 + \cos(n\phi - \phi_0))$, with units: nm, kJ/mol, rad.}
    \begin{tabular}{llll}
        \toprule
         \toprule
         \bf Bonds & $r_0$ & $k_r$ & $D$ \\
        \midrule
        \ce{H2} & 0.074451338 & 352643.84 & 450.529259 \\
        \ce{OH} & 0.097519183 & 472095.947 & 441.538052 \\
         \ce{O2} & 0.119614093 & 839237.053 & 529.522227 \\
         \ce{H2O} & 0.096205047 & 511168.368 & 476.339341 \\
         \ce{HO2} & 0.131276144 & 462105.576 & 306.492086 \\
          & 0.09757723 & 451086.258 & 441.538052 \\
         \ce{H2O2} & 0.143811214 & 316472.666 & 234.832386 \\
          & 0.096527728 & 493591.819 & 441.538052 \\
         \midrule
         \midrule
         \bf Angles & $\theta_0$ & $k_\theta$ \\
         \midrule
         \ce{H2O} & 1.83330493 & 448.270537 \\
         \ce{HO2} & 1.84308154 & 698.279951 \\
         \ce{H2O2} & 1.73955327 & 638.30292 \\
         \midrule
         \midrule
         \bf Dihedrals & $\phi_0$ & $k_\phi$ & $n$ \\
         \midrule
         \ce{H2O2} & 0.0 & 5.9853054 & 1 \\
          & 3.14159265 & -3.0246061 & 2 \\
          & 0.0 & -2.185907 & 3 \\
          & 3.14159265 & 0.650878 & 4 \\
         \bottomrule
         \bottomrule
     \end{tabular}
     \label{tab:ff}
 \end{table}

\noindent
\textbf{Molecular dynamics simulations}. All simulations were done using the Atomic Simulation Environment (ASE).\cite{ase_2017} Langevin dynamics with a friction parameter of 2 ps$^{-1}$ and 0.5 fs time steps were used for the MD simulations of aspirin and water. For the free energy surfaces for the hydrogen combustion reactions we used well-tempered metadynamics at 300 K using the Plumed plugin\cite{plumed_2014}. A Gaussian potential of 5 kJ/mol height and 0.05 width along the two collective variables was deposited every 100 steps for the hydrogen combustion simulations, coupled with the a Langevin thermostat with a friction coefficient of 0.2 ps$^{-1}$ and 0.2 fs time steps, consistent with the reference.\cite{Guan2023}

\section{Data availability}
\noindent
Coordinates of geometries, energy and forces for hydrogen combustion are found in the original dataset\cite{Guan2022} and is available at https://doi.org/10.6084/m9.figshare.19601689. IRC dilation data and active learning generated data\cite{Guan2023} used in the training are available at 
https://doi.org/10.6084/m9.figshare.23290115.v1. 

Source data for Figures 1-3 is available with this manuscript. 

\section{Code availability}
\noindent
The GitHub repository for NewtonNet is publicly available and open source at https://github.com/THGLab/NewtonNet. We also designed a command line interface to facilitate faster implementation for non-programmers. 

\section{Acknowledgment}
\noindent
E.C.-Y.Y. and T.H-G. thank the CPIMS program, Office of Science, Office of Basic Energy Sciences, Chemical Sciences Division of the U.S. Department of Energy under Contract DE-AC02-05CH11231 for support. This work used computational resources provided by the National Energy Research Scientific Computing Center (NERSC), a U.S. Department of Energy Office of Science User Facility operated under Contract DE-AC02-05CH11231, and the Lawrencium computational cluster resource provided by the IT Division at the Lawrence Berkeley National Laboratory (Supported by the Director, Office of Science, Office of Basic Energy Sciences, of the U.S. Department of Energy under Contract No. DE-AC02-05CH11231).

\section{Author contributions}
\noindent
E.C.-Y.Y. and T.H.G. designed the project. E.C.-Y.Y. carried out all training and implemented and executed all workflows. E.C.-Y.Y. and T.H.G. discussed the results and wrote the manuscript.

\section{Competing interests}
\noindent
The authors declare no competing interests.

\bibliography{references}

\bibliographystyle{unsrt}

\end{document}


\author{Eric C.-Y. Yuan$^{1,4}$ and Teresa Head-Gordon$^{1-4}$}
\date{}
\maketitle

\begin{center}
\vspace{-24pt}
$^1$Kenneth S. Pitzer Theory Center and Department of Chemistry, $^2$Department of Bioengineering, $^3$Department of Chemical and Biomolecular Engineering,  University of California, Berkeley, CA, 94720 USA\\
$^4$Chemical Sciences Division, Lawrence Berkeley National Laboratory, Berkeley, CA, 94720 USA

corresponding author: thg@berkeley.edu
\end{center}

\section{Supplementary Figures}

\begin{figure}[H]
\centering
\includegraphics[width=0.9\textwidth]{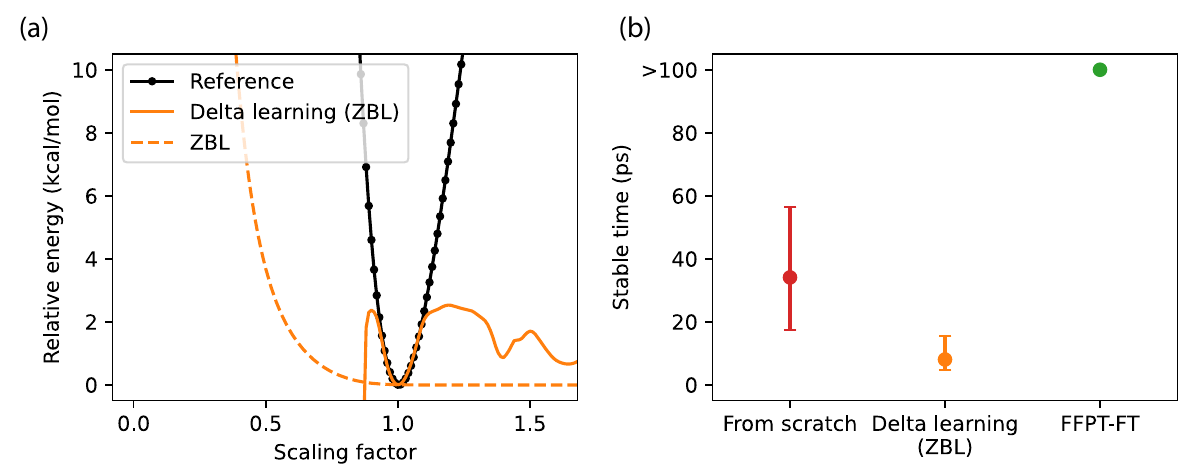}
    \caption{\textit{Delta learning from the ZBL baseline.} The Ziegler-Biersack-Littmark (ZBL) potential has been proposed to handle the collision behavior of atoms and improve the simulation stability, but we show that this is an ineffective approach since it is designed for extremely short range repulsion. Indeed, both (a) the PES profile and (b) MD stability show no sign of improvement over the vanilla model.}
    \label{fig:zbl}
\end{figure}

\begin{figure}[H]
\centering
\includegraphics[width=0.9\textwidth]{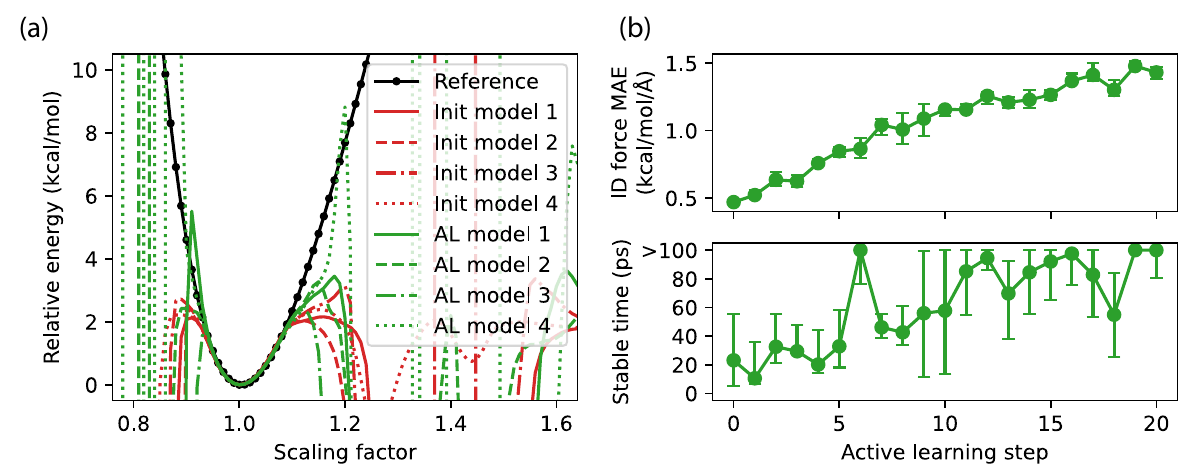}
    \caption{\textit{Active learning.} The active learning approach has been widely adopted to improve simulation stability, but here we see that it is an ineffective approach since it not only requires expensive computational resources, but also dilutes the in distribution (ID) data. Indeed, (a) the PES profile shows little improvement over the initial models after many rounds of active learning, and (b) the ID error increases significantly with statistically limited simulation stability improvement.}
    \label{fig:zbl}
\end{figure}

\begin{figure}[H]
\centering
\includegraphics[width=0.9\textwidth]{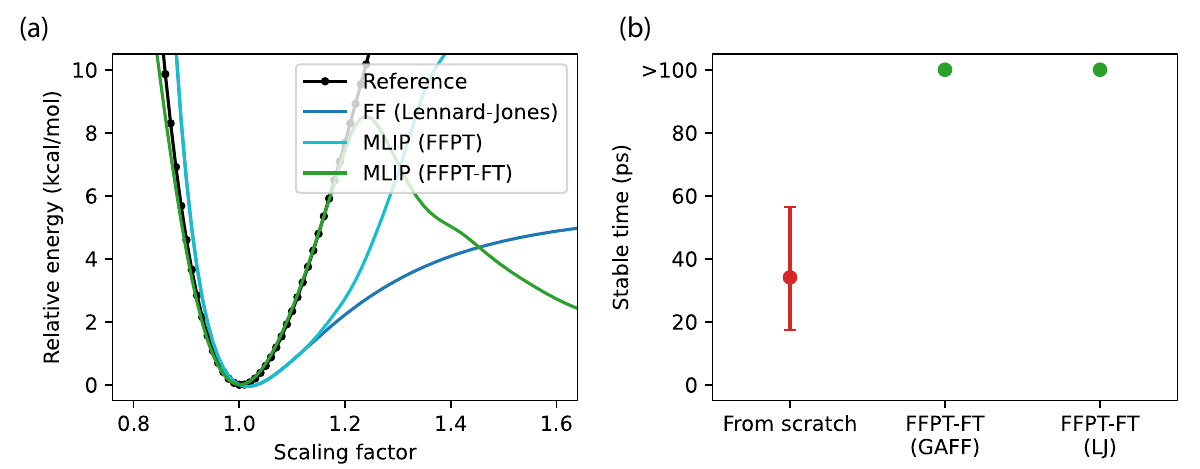}
    \caption{\textit{Effective pre-training using a Lennard-Jones potential.} Even with a simpler Lennard-Jones potential to describe the aspirin molecule, our FFPT-FT approach is still effective in both (a) PES profile and (b) MD stability. Note that, compared to the harmonic potentials like GAFF, the large force magnitude from the LJ potential requires a smaller 0.1 \r{A} rattling, as well as a Huber loss instead of a mean squared loss. However the central idea of our method still holds well using this simpler FF.}
    \label{fig:lj}
\end{figure}

\begin{figure}[H]
\centering
\includegraphics[width=0.9\textwidth]{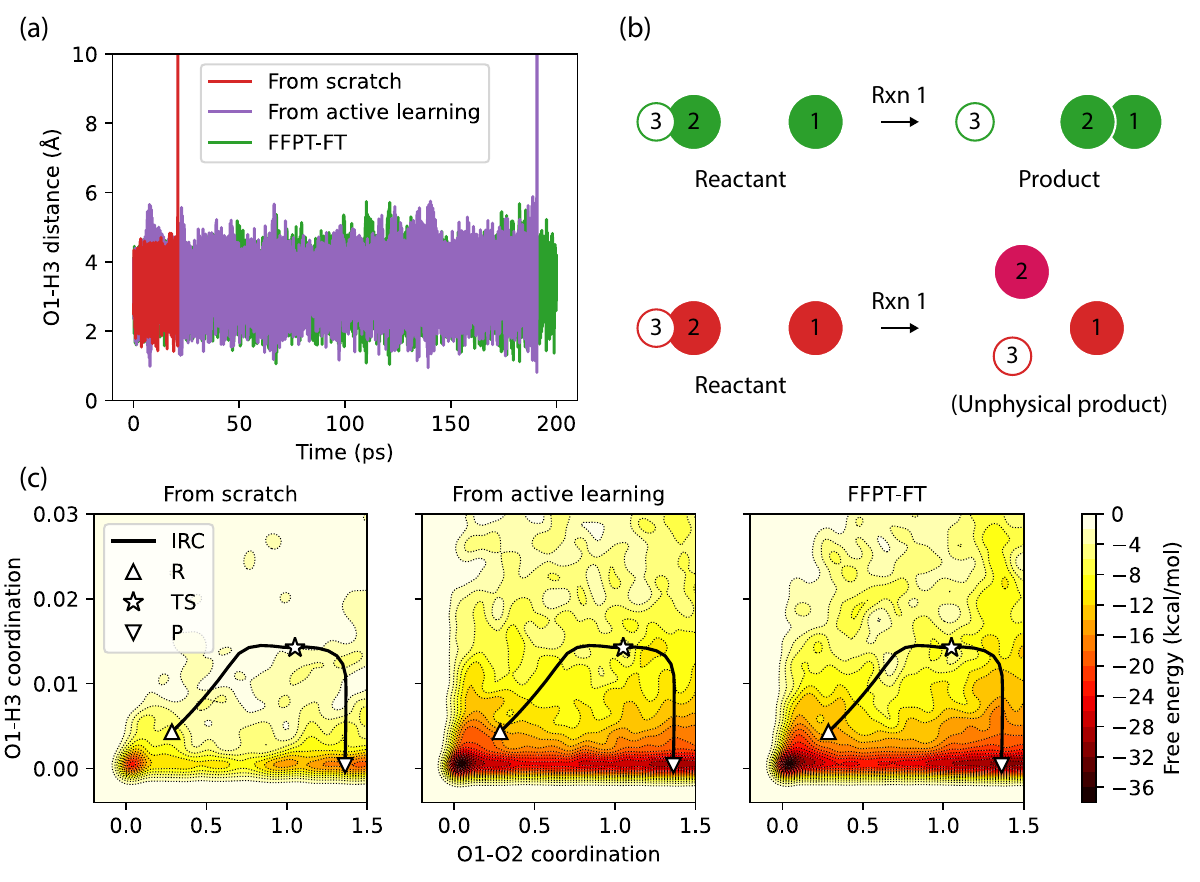}
    \caption{\textit{Hydrogen combustion reactions improved by FFPT-FT for reaction 1.} (a) The MLIP trained from scratch showed catastrophic OOD behaviors in simulation (red) and even the active learning model fails at the end of the trajectory (purple). The FFPT-FT model is stable over the entire trajectory (green). (b) The non-smooth PES leads to simulation failure by predicting unphysical products when running metadynamics with the MLIP from scratch (red) whereas the MD is stable using the FFPT-FT model (green) that yields correct products. (c) The unstable simulation trajectory of the MLIP from scratch reconstructs a free energy surface (FES) that does not resemble the ground truth, with an overstabilized product state with a large entropy component to the free energy. While improved with active learning, the FFPT-FT shows easy converging of the FES surface.}
    \label{fig:h2comb2}
\end{figure}

\begin{figure}[H]
\centering
\includegraphics[width=0.9\textwidth]{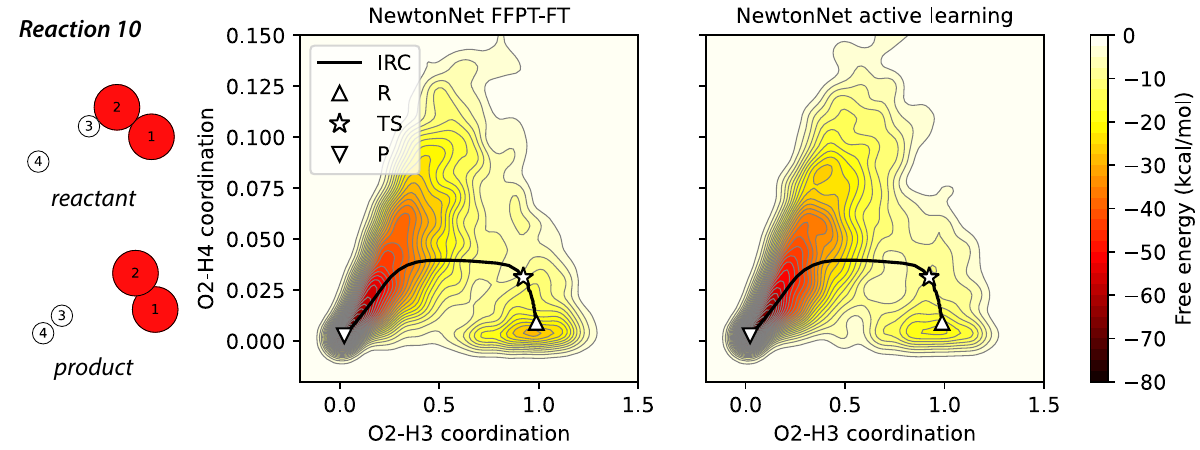}
\includegraphics[width=0.9\textwidth]{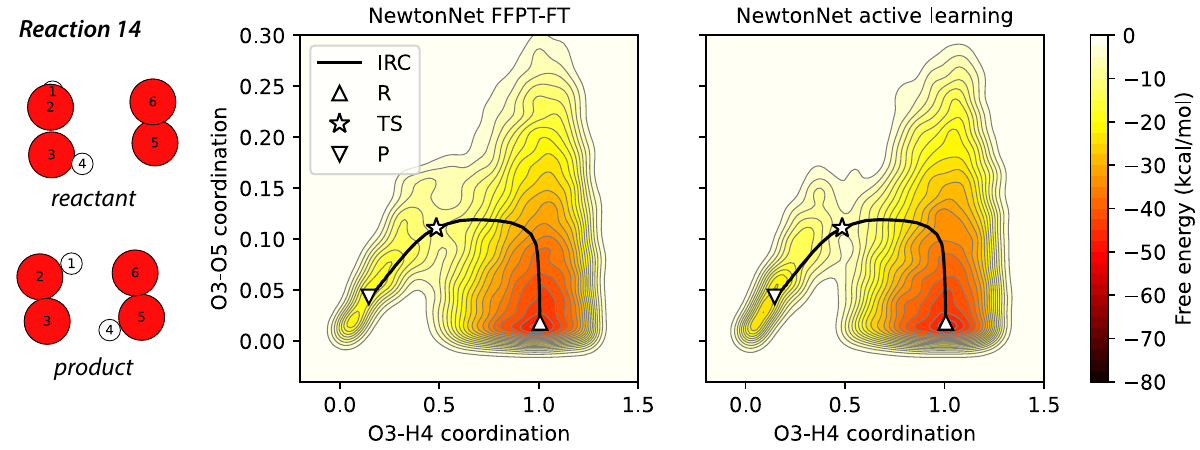}
\includegraphics[width=0.9\textwidth]{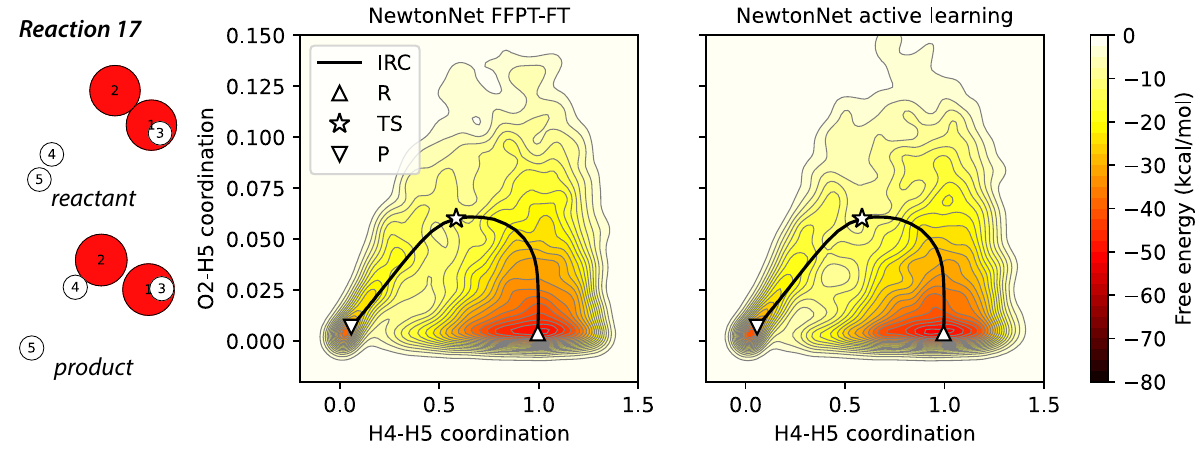}
\includegraphics[width=0.9\textwidth]{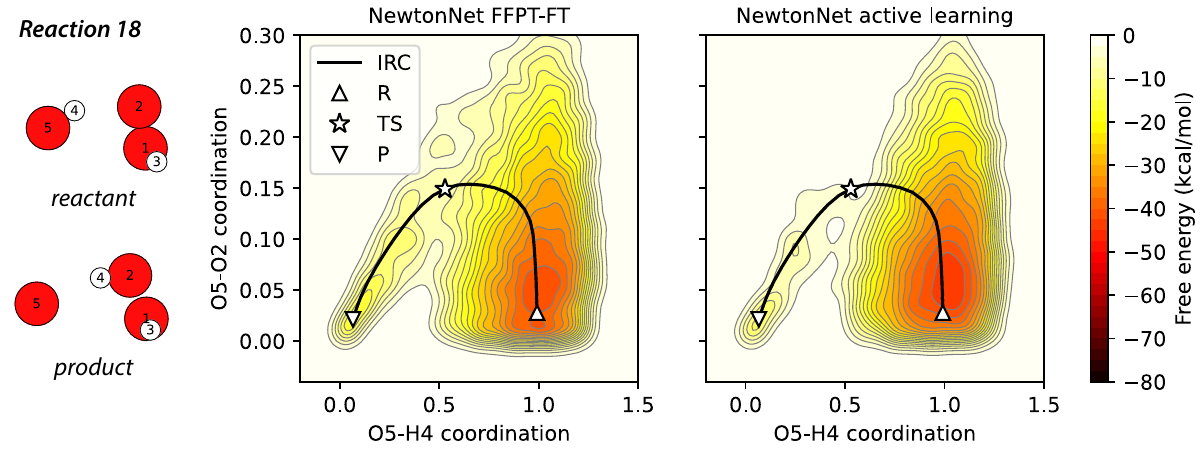}
    \caption{\textit{Hydrogen combustion reactions comparing active learning with FFPT-FT.} (a) Reaction 10 (b) Reaction 14 (c) Reaction 17 (d) Reaction 18.}
    \label{fig:h2comb2}
\end{figure}

\begin{figure}[h!]
\centering
\includegraphics[width=0.95\textwidth]{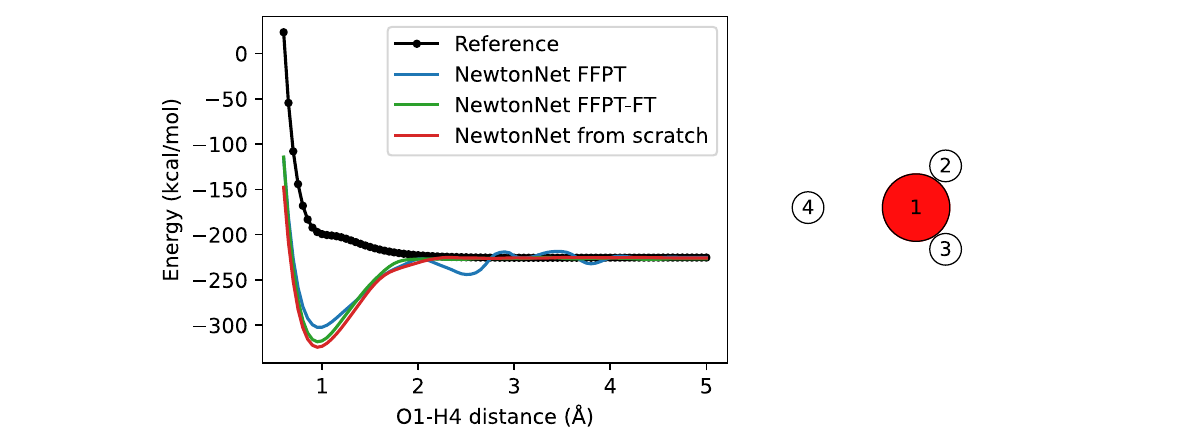}
    \caption{\textit{The potential energy profile of hydronium.} While the force-field pre-training includes unphysical conformations of many chemical species, it doesn't sample high energy chemical species like the hydronium ion for hydrogen combustion. Therefore, the force-field pre-training does not solve the hydronium problem that we found in our previous work.}
    \label{fig:hydronium}
\end{figure}

\newpage

\section{Supplementary Tables}

\begin{table}[htbp]
\begin{center}
\begin{tabular}{ | m{6cm} | m{1cm}| m{1cm} | m{1cm} |}
  \hline 
  No. Reaction & Atoms & $\mathrm{DoF}$ & $\mathrm{DoF_{int}}$ \\ 
  \hline
  \textbf{Association/Dissociation} & & & \\
  5. H$_2\longrightarrow \;$2H & 2 & 6 & 1 \\
  6. O$_2\longrightarrow \;$2O & 2 & 6 & 1 \\
  7. OH$\;\longrightarrow \;$O+H & 2 & 6 & 1\\
  8. H+OH $\longrightarrow$ H$_2$O & 3 & 9 &  3 \\
  9. H+O$_2 \longrightarrow \;$HO$_2$ & 3 & 9 & 3 \\
  15. H$_2$O$_2$ $\longrightarrow\;$2OH & 4 & 12 & 6 \\  
  \hline
  \textbf{Substitution} &&&\\
  16. H$_2$O$_2$+H $\longrightarrow\;$H$_2$O+OH & 5 & 15 & 9 \\
  \hline
  \textbf{O-transfer} &&&\\
  1.  OH+O $\longrightarrow\;$ H+O$_2$ & 3 & 9 & 3 \\
  11. HO$_2$+H$\;\longrightarrow\;$2OH & 4 & 12 &  6 \\
  12. HO$_2$+O$\;\longrightarrow\;$OH+O$_2$ & 4 & 12 &  6 \\ 
  \hline
  \textbf{H-transfer} &&&\\
  2. O+H$_2 \longrightarrow\;$OH+H& 3 & 9 & 3 \\
  3. H$_2$+OH $\longrightarrow\;$H$_2$O+H & 4 & 12 & 6\\
  4. H$_2$O $\longrightarrow\;$2OH & 4 & 12 & 6 \\
  10. HO$_2$+H $\longrightarrow\;$H$_2$+O$_2$ & 4 & 12 & 6 \\
  13. HO$_2$+OH $\longrightarrow\;$H$_2$O+O$_2$& 5 & 12 & 9 \\
  14. 2HO$_2\longrightarrow\;$ H$_2$O$_2$+O$_2$& 6 & 18 & 12 \\
  17. H$_2$O$_2$+H $\longrightarrow\;$HO$_2$+H$_2$& 5 & 15 & 9 \\
  18. H$_2$O$_2$+O $\longrightarrow\;$ HO$_2$+OH & 5 & 15 & 9 \\
  19. H$_2$O$_2$+OH $\longrightarrow\;$H$_2$O+HO$_2$& 6 & 18 & 12 \\
  \hline
\end{tabular}
\end{center}
\caption[The 19 reactions contained in the hydrogen combustion benchmark dataset.]{\textit{The 19 reactions contained in the hydrogen combustion benchmark dataset.} The number of atoms involved in each reaction, the total number of degrees of freedom ($\mathrm{DoF}$) in Cartesian coordinates, and total number of degrees of freedom in ICs ( $\mathrm{DoF}_{int}$.)}
\label{tb:reactions}
\end{table}